\documentclass{IEEEcsmag}

\usepackage[colorlinks,urlcolor=blue,linkcolor=blue,citecolor=blue]{hyperref}
\expandafter\def\expandafter\UrlBreaks\expandafter{\UrlBreaks\do\/\do\*\do\-\do\~\do\'\do\"\do\-}
\usepackage{upmath,color}
\usepackage{color,soul}

\jvol{XX}
\jnum{XX}
\paper{8}
\jmonth{Month}
\jname{IEEE Security \& Privacy}
\jtitle{IEEE Security \& Privacy}
\pubyear{2026}

\begin{document}

\sptitle{Article Type: Department  (Operational Technology)}

\title{Building an Open Source Operational Technology Pentesting Platform: \\ Lessons from LINICS}

\author{Awais Rashid}
\affil{Hacktonics Ltd, United Kingdom}

\author{Joseph Gardiner}
\affil{Hacktonics Ltd, United Kingdom}

\author{{L}ouise Evans}
\affil{Hacktonics Ltd, United Kingdom}

\markboth{DEPARTMENT}{DEPARTMENT}

\begin{abstract}
Information Technology (IT) security professionals have ready access to open-source platforms such as Kali Linux. But no such platform exists for Operational Technology (OT) that underpins Industrial Control Systems. We discuss experiences of architecting, building and releasing LINICS, an open-source platform for OT pentesting and security analysis.
\end{abstract}

\maketitle

\chapteri{O}perational Technology (OT) underpins Industrial Control Systems (ICS) driving critical infrastructures on which we rely everyday: those delivering water and power to our homes, running manufacturing environments, and ensuring safe, continued operation of transportation systems. While there is usage of IT, especially in office-based settings and computing systems to run SCADA (Supervisory Control and Data Acquisition) software, the cyber-physical systems driving production environments depend on OT. This includes specialist devices such as Programmable Logic Controllers (PLCs) and Remote Terminal Units (RTUs) utilising proprietary hardware and networking protocols, Human-Machine Interfaces (HMIs) built on custom or legacy embedded operating systems, and vendor-specific software for programming, management and control.

The security issues in OT, e.g., legacy devices and protocols, lack of state-of-the-art hardware and software security mechanisms, and safety and reliability implications of patching and upgrades are well-documented (see, e.g.,~\cite{gupta24}). While some standard IT security tools are pertinent to pentesting OT, specialist tools -- attuned to the proprietary nature of OT devices, architectures and networking protocols -- are needed in order to probe and analyse vulnerabilities and understand the potential safety and reliability impacts.

\begin{figure*}[t]
\centerline{\includegraphics[width=\textwidth]{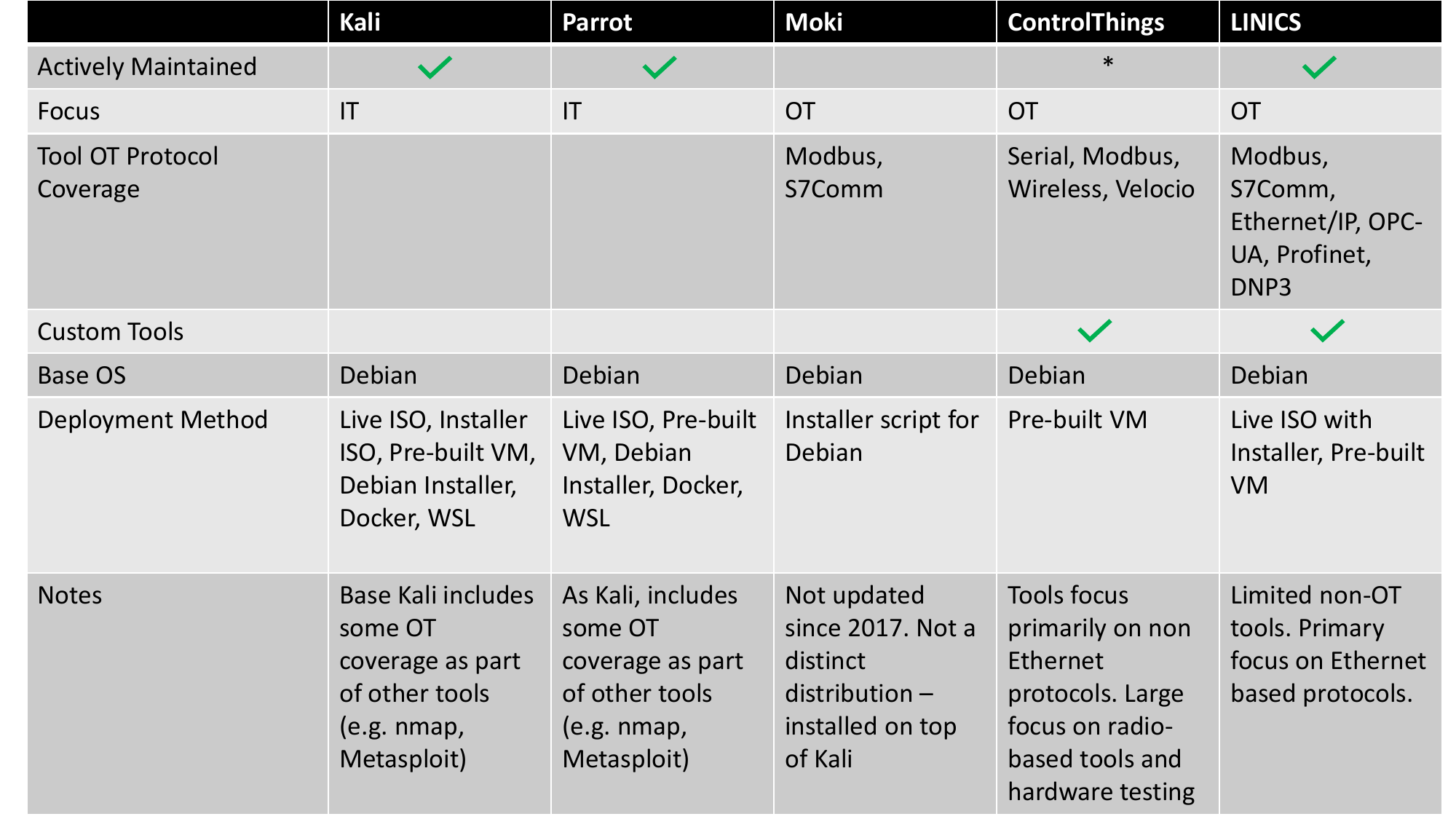}}
\caption{Comparison of OT testing Linux distributions. \textsuperscript{*}ControlThings has not been updated since 2022, including updates to the individual tools on GitHub}
\label{fig:tools-comparison}
\end{figure*}

IT security professionals have ready access to tools, e.g., Kali Linux (\url{https://www.kali.org/}). Similar resources are lacking for OT. The typical approach has been for OT pentesters to roll their own\textemdash often as a variant of Kali or similar platforms. There is no single baseline platform that provides relevant tools systematically mapped to a de facto industry framework, e.g., MITRE ATT\&CK\textsuperscript{\textregistered} for ICS (\url{https://attack.mitre.org/matrics/ics}). Nor is there a singular collection of the array of OT security analysis tools that those transitioning to a career in OT security can readily access and learn. The problem is compounded as many such tools are legacy tools themselves -- often not maintained for a long period -- hence posing additional overheads and complexity to users.

\section{PREVIOUS INITIATIVES}

There have been two previous initiatives aimed at building OT-focused Linux\textsuperscript{\textregistered}\footnote{Linux\textsuperscript{\textregistered} is the registered trademark of Linux Torvalds in the U.S. and other countries.} distributions: Moki (\url{https://github.com/moki-ics/moki}) and ControlThings (\url{https://www.controlthings.io/}). However, neither provides an extensive set of OT security tools nor have they been maintained or updated for several years. Figure~\ref{fig:tools-comparison} provides a comparison of LINICS with existing platforms, including IT-focused distributions, Kali and ParrotOS, which can have OT tools added to them manually, but feature a number of IT-focused tools that may be risky to use against OT devices.

Note that, while tools such as MITRE Caldera (and its specific variant for OT) exist (\url{https://www.mitre.org/resources/caldera-ot}), and are mapped to MITRE ATT\&CK\textsuperscript{\textregistered}, their focus is on adversary emulation. They can be used to generate data to train OT security teams and are, therefore, complementary to efforts such as LINICS, which is focused on providing a security and penetration testing platform rather than adversary emulation capabilities. 

\section{DESIGN REQUIREMENTS}

When we set out to develop LINICS (\textbf{Lin}ux\textsuperscript{\textregistered} for \textbf{ICS})(\url{https://linics.org/}), the analysis in Figure~\ref{fig:tools-comparison} formed the basis for the requirements that guided our design choices:

\begin{itemize}
\item[{\ieeeguilsinglright}] {\it Not just some tools added onto Kali}---We, ourselves, are regular users of Kali; it is an invaluable resource for the security community. However, while some IT security tools are pertinent to OT (and several are included in LINICS), not all are required. Our experience of rolling our own Kali variants for many years highlighted -- echoed by informal feedback from other OT security professionals -- a need for a similar platform specifically focused on OT with dedicated tools clearly mapped. Furthermore, and especially for newcomers to OT security, Kali variants lead to confusion as to which tools they ought to use, for what purpose and at what stage of an OT test. 

\item[{\ieeeguilsinglright}] {\it Incorporation of legacy OT security tools}---A number of OT security tools have been developed by individuals or small teams and many are no longer actively maintained. However, they remain a useful resource especially for pentesting legacy OT devices and protocols. The platform needed a systematic approach to incorporating and updating such tools.

\item[{\ieeeguilsinglright}] {\it Ease of installation and use}---Good user experience was a key motivation, whether installing from scratch, utilising the pre-built Virtual Machines or using LINICS in pentesting tasks.

\item[{\ieeeguilsinglright}] {\it Supporting newcomers to OT security}---Often OT security professionals pivot from previous roles as safety engineers. Alternatively, those engineers wish to upskill and understand safety implications of vulnerabilities. Even when IT security professionals transition into OT security, they need clear signposting through the platform. This underpinned our decision to map all tools to MITRE ATT\&CK\textsuperscript{\textregistered} for ICS. 

\item[{\ieeeguilsinglright}] {\it Maintainability and user support}---We observed that the two previous dedicated OT-focused distributions had not been actively maintained. Longer term maintainability and user support was, therefore, not only a technical consideration but also needed to align with our company's medium to long term strategic plans. 

\end{itemize}

\section{LESSONS LEARNT}

LINICS (version 1.01 at the time of this article) is based on Debian Live 12.8 (Bookworm). We chose Debian for its stability, its usage as the base for other security-focused distributions such as Kali and Parrot OS, and our longstanding experience in using and customising various Debian-based distributions. We started with Debian Live 12.8 as our base, and used Cubic (\url{https://github.com/PJ-Singh-001/Cubic}) and bespoke scripts for our pipeline to customise it with various tools and configurations for LINICS. We selected Gnome as the desktop environment due to our own familiarity as users and our knowledge of customising it. 
We now discuss the lessons learnt.

\begin{figure*}[!ht]
\centerline{\includegraphics[width=\textwidth]{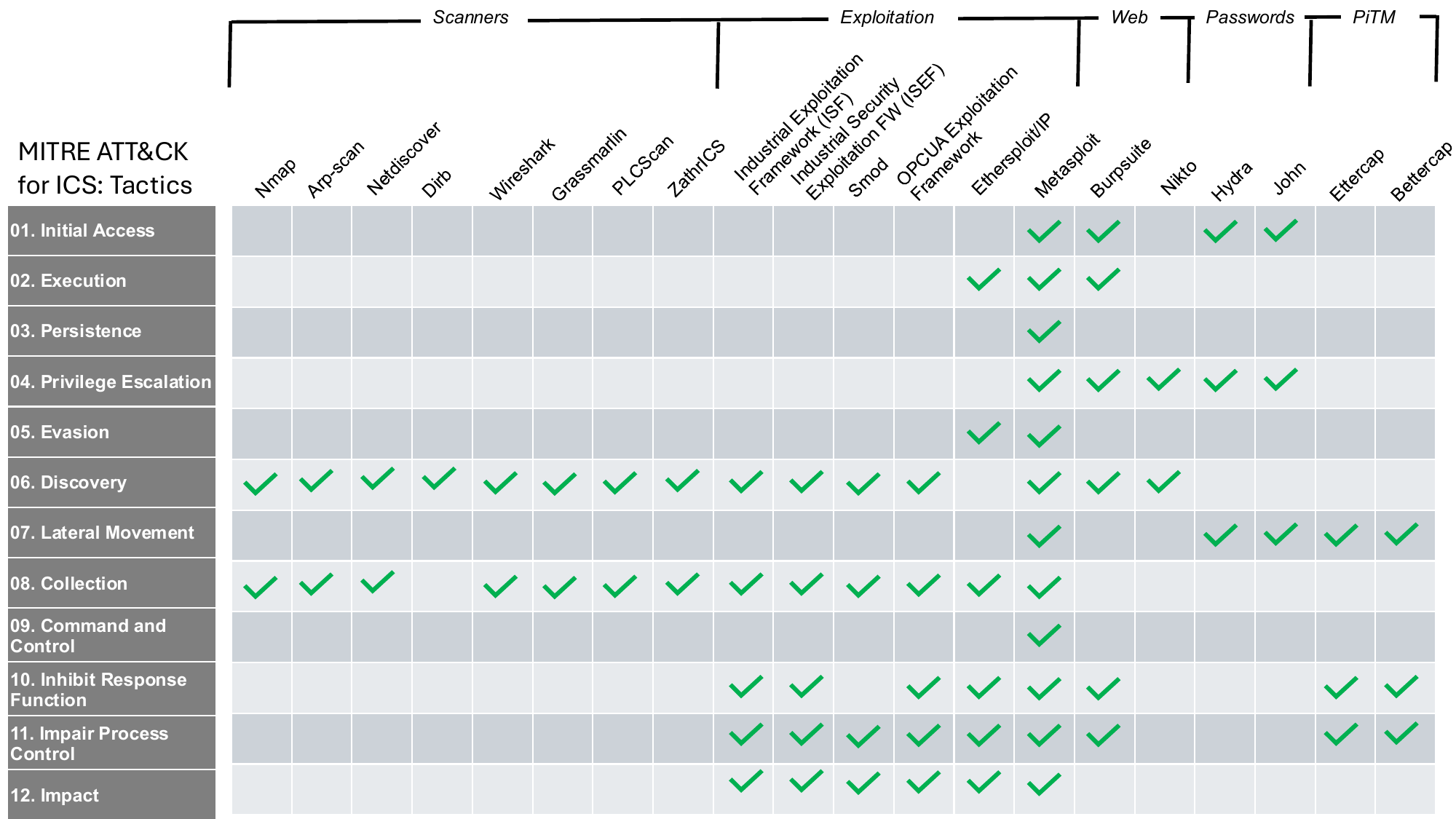}}
\caption{Mapping of tools included in LINICS 1.01 to MITRE ATT\&CK\textsuperscript{\textregistered} for ICS}
\label{fig:mitre-mapping}
\end{figure*}

\subsection{Lesson 1: State of OT Security Tools is Patchy}\label{sec:lesson1}

LINICS includes various categories of tools, which are all mapped to MITRE ATT\&CK\textsuperscript{\textregistered} for ICS (Figure~\ref{fig:mitre-mapping}) and these mappings are, in turn, setup as part of the Gnome Application Menu so that a user can easily find tools for each ATT\&CK stage. Tools include: {\it Scanners} (both general purpose and specialist OT-focused); {\it Exploitation Tools} (OT-focused frameworks and ICS-specific modules within Metasploit); {\it Web Security Tools} (many OT devices incorporate embedded web servers for remote configuration and interaction); {\it Password Crackers} (to proble password vulnerabilities); and {\it Person-in-the-Middle} (many ICS protocols are susceptible). Though not mapped to MITRE ATT\&CK, the Volatility framework, a general purpose forensics framework for extraction of artefacts from RAM, is also included. 
  
The mapping in Figure~\ref{fig:mitre-mapping} shows that the state-of-the-art with regards to OT security tooling has significant gaps. While some of the ATT\&CK stages are well-covered, e.g., Discovery and Collection, others have partial coverage. Most significantly, there is need for additional tooling for stages such as Execution, Persistence, Evasion, and Command \& Control. At present, primarily the coverage here comes from the broad applicability of Metasploit and some of the ICS-specific modules that we have included within Metasploit. The same is true of Lateral Movement, where there are no OT-specific tools. We also noted a general lack of readily available ICS forensics tools.

Figure~\ref{fig:mitre-mapping} provides a blueprint for the LINICS core development team, and OT security researchers and practitioners in general, to proactively focus on developing tools that address this lack of coverage (discussed further in the roadmap). 

\subsection{Lesson 2: Modular Architectures Minimise Downstream Costs, especially for Managing Legacy Tools}\label{sec:lesson2}

We containerised (using docker) a number of legacy applications and python scripts useful for OT pentesting, namely ISF, ISEF, PLCScan and SMOD. In other cases, where docker images were available, or simply for consistency of user experience, we respectively used the available docker images (Nikto and Bettercap) or containerised the frameworks (Volatility and OPCUA-EF). This enabled us to, on the one hand, address dependency clashes amongst the various Python tools and cater to their specific Python versions, and on the other, minimise bloat by packaging non-clashing tools within a single image.

Another key motivator was update and change management. The docker images use a uniform architecture whereby any updates require minimal changes to mitigate against any ripple effects. We recognise that there may be a small performance overhead due to containerisation. However, on modern multi-core processors, this isn't significant and is outweighed by the maintainability and dependency management benefits. This is further evidenced by the minimal resource requirements for LINICS: 4 GB RAM, 30 GB hard disk and 2 processor cores, which are serviceable by most computing hardware. Our tests show that LINICS works on 2 GB RAM and 1 Processor Core, however, performance takes a serious hit.

\subsection{Lesson 3: Usability Begins from Installation}\label{sec:lesson3}

Recent work has extended the classical notion of usability that focused on end-users of security technologies~\cite{adams1999, whitten1999, cybok-hf} to security usability considerations for professionals, such as software developers, e.g.,~\cite{acar2017, rashid2021}, incident responders, e.g.,~\cite{stevens2022} and OT professionals~\cite{li2024}. We extend this further to OT pentesters, especially as most newcomers to the field are either transitioning from IT security to OT security and hence would have misaligned mental models of security needs of OT systems. Alternatively, they are safety engineers moving to OT security and often do not have familiarity with pentesting distributions such as Kali and, in many cases, with shell commands and code. Li et al.~\cite{li2024} have highlighted the challenges and complexity of configuring security settings on PLCs. Usability and user experience (UX) was, therefore, a key consideration, addressed across multiple dimensions:

\begin{itemize}

\item \emph{Installation:} Doing away with the default Debian installer and configuring the more user-friendly Calamares installer within Debian Live to ensure that the LINICS image could be installed and configured with ease. We clearly documented any problems and solutions in the Release Notes (available on Github) to ensure that users have ready access to \emph{known issues} and how to troubleshoot them.

\item \emph{Pre-built resources:} VM images for major hypervisors.
  
\item \emph{Consistent user experience:} Uniform shell launchers for all command line tools, associated man pages, alongwith graphical launchers and custom, informative, icons for the application launch menu using the ATT\&CK mapping.
  
\item \emph{Guides and documentation:} \emph{Getting started with LINICS} guides providing step-by-step guidance on: installation, various tools within LINICS and step through use cases with specific tools.
  
\item \emph{Alleviating dependency-management workload:} Containerisation of legacy and non-legacy tools, so that users can focus on their primary task: OT pentesting.

\end{itemize}

\subsection{Lesson 4: Lowering the Barriers to Entry into OT Security is Critical}\label{sec:lesson4}

Our experience is that newcomers to the field, especially those pivoting from safety engineering or network engineering backgrounds, may not be regular users of Linux\textsuperscript{\textregistered} based platforms. However, these professionals have substantial experience in working with OT devices and OT networks. Making their transition to OT security smoother is important in addressing the workforce gaps in this domain. This was a key motivator behind development of the ZathrICS scanner which combines a number of scanning features into an easy to use point-and-click interface for active asset discovery and device enumeration. In addition, we also set up an easy to use \emph{LINICS Launcher} which categorises the tools based on their functionality (e.g., scanners, exploitation, etc.) and provides information on each tool along with basic usage instructions (similar to what may be found in a man page) in one place. This provides newcomers with a one-stop resource to familiarise themselves with the purpose of the tools, how to use them and launch them directly.

\begin{figure*}[!ht]
\centerline{\includegraphics[width=\textwidth]{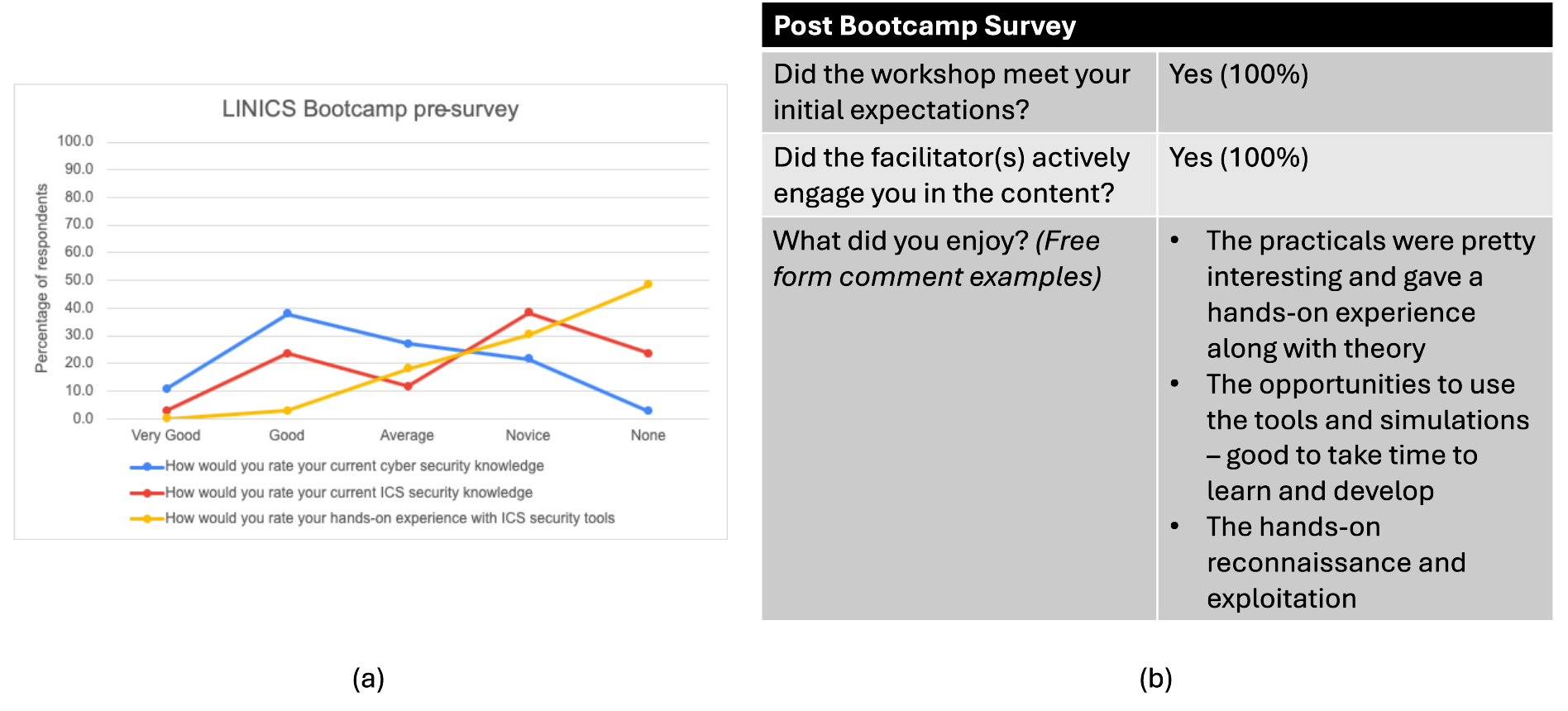}}
\caption{Feedback from bootcamp attendees (27 respondents): (a) Pre-bootcamp Survey (b) Post-bootcamp Questionnaire}
\label{fig:feedback}
\end{figure*}

\subsection{Lesson 5: If you Build it, they will Come ... But you Must Maintain it!}\label{sec:lesson5}

Building and releasing a platform is just the first step. Supporting users and longer term maintenance and update is also critical. We already highlighted some of our architectural choices above to ensure modular updates and minimise knock-on effects. As a small startup-based team, we also needed to consider how best to support users whilst ensuring that LINICS remains an active and long term resource. For the former, we set up easy to use resources: ISO, pre-built VMs, release notes, issue reporting via Github and a Discord server for users to engage with us and each other. For the latter, we aligned LINICS with our company's long term strategy. All our training was transitioned from our custom Kali setups to LINICS so that it is embedded at the core of current and future training products and services being developed. We also established LINICS as a core strand within our R\&D roadmap with a focus on adding new tools, and maintaining/updating legacy ones incorporated into LINICS, to ensure that the gaps in the ATT\&CK mapping (Figure~\ref{fig:mitre-mapping}) are plugged alongside development of new training products and services for our client base.

However, the current release does not have an apt (or similar) repository for users to directly install or update tools. This was driven by limited development resources and condensed timescales. We, therefore, prioritised tool coverage, testing and user experience. This is already a \emph{pain point} from a maintenance standpoint, which we plan to address in an upcoming release. 

\section{USER EXPERIENCE AND FEEDBACK}

We ran four free-to-attend bootcamps -- the first public release and usage of LINICS -- across the South West of England during March 2025 with 46 participants. Attendees came from a range of industries including: defence, cybersecurity, healthcare, energy and higher education (both academics and students).
The bootcamps were publicised as aimed at newcomers to OT security. Therefore, the topics covered included background on ICS security problems, the challenges of securing legacy devices and platforms, and the safety, reliability and real-time impacts of security violations.

Pre-bootcamp (Figure~\ref{fig:feedback}(a)), most attendees rated their cybersecurity knowledge as good or average, their knowledge of ICS security as good, novice or none, and little to no hands-on-experience with OT security tools. Post-bootcamp (Figure~\ref{fig:feedback}(b)), all respondents were uniformly positive, highlighting the practical and hands-on experience with the tools. Several noted the need for more hands-on ICS pentesting as their priority.

\begin{figure*}[!ht]
\centerline{\includegraphics[width=0.75\textwidth]{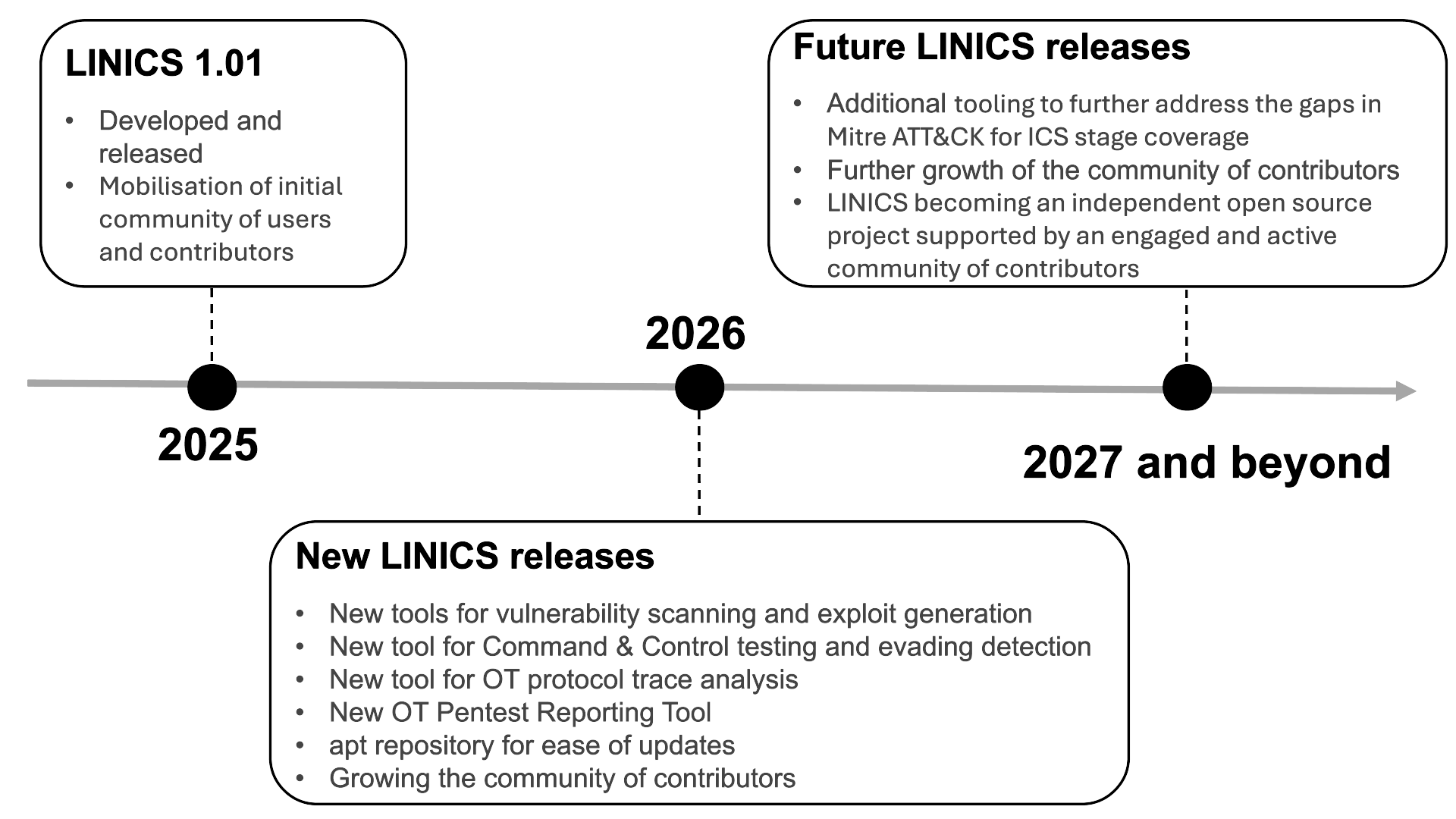}}
\caption{Roadmap for LINICS}
\label{fig:roadmap}
\end{figure*}

Beyond the bootcamps, LINICS was used extensively at a dedicated OT focused capture-the-flag (CTF) event held at the University of Bristol in summer 2025. Many of the 42 participants (mainly from industry) used LINICS as their main attack machine. LINICS is also used in the University of Bristol's MSc in Infrastructures Security programme for the 2025-26 academic year replacing a modified Kali instance. We have also been informed by multiple other universities of plans to incorporate it into their teaching and research. 

\section{SAFE USE GUIDANCE}\label{sec:safe}

As the tools included in LINICS work against real-world devices, protocols and systems, we offer the following safe use guidelines. 

\begin{itemize}

\item Users should be aware of the risks to live systems when pentesting. Irresponsible utilisation of tools against production systems can lead to disruption or outages. Training and experience on test setups is critical before conducting pentesting on production environments. 
  
\item For those transitioning from IT security, traditional IT pentesting techniques do not translate readily. Unfettered network scans can and do interfere with legacy devices. Tools, such as ZathrICS, are provided within LINICS to reduce the risk arising from such active scans. 

\item When testing production environments, clearly establish the scope and how far a pentest is allowed to proceed within the system. Establish checkpoints as part of the scoping and seek confirmation before proceeding after a checkpoint is reached. Often reaching a key device or service is sufficient evidence of security vulnerabilities. Actual manipulation serves no real purpose and can risk disruption to production systems. Impacts can always be demonstrated on test setups if needed.

\end{itemize}

\section{ROADMAP}\label{sec:roadmap}

The mapping to MITRE ATT\&CK\textsuperscript{\textregistered} for ICS (Figure ~\ref{fig:mitre-mapping}) highlighted a number of tooling gaps particularly with regards to \emph{Execution, Persistence, Evasion} and \emph{Command \& Control}. Our analysis also identified a lack of OT-focused forensics tools. These insights have played a key part in informing our immediate and future roadmap (Figure~\ref{fig:roadmap}) to plug the gaps through:

\begin{itemize}

\item OT vulnerability scanning and exploit generation tools \emph{[Stages: Execution; Persistence]};
\item OPCUA-based Command \& Control, including stealthy execution of such actions \emph{[Stages: Evasion; Command \& Control]};
\item OT protocol trace analyser, to add dedicated OT protocol analysis capability for \emph{Forensics}.

\end{itemize}

These tools are being developed both by the core development team and through co-supervised student projects in collaboration with academia. The roadmap is further informed by the lessons learned through our experience, e.g., the incorporation of an apt repository for ease of updates and an OT pentest reporting tool based on feedback from bootcamps regarding more hands-on OT pentesting experience.

\section{CONCLUSION}

The conception of LINICS was borne out of a frustration shared with many other OT security professionals of having to roll our own variants of Kali or similar platforms. Our experience shows that such a platform must go beyond being a mere collection of tools and must pay close attention to issues of legacy tooling, user experience, and long-term maintenance. We learnt other key lessons along the way. For instance, all software projects, however well-intentioned, are resource constrained. This resulted in some gaps in our maintenance infrastructure for LINICS which we plan to fill in the future.

The user feedback from bootcamps showcased the value of hands-on training with a platform such as LINICS and also highlighted the importance of further OT pentesting tools and resources. Our intention is that LINICS will be driven by community need rather than a single actor's priorities (including ours) and form the nucleus around which a community effort will mobilise.

\section{ACKNOWLEDGMENTS}
This work was supported by the Department of Science Innovation and Technology and InnovateUK under the grant: Bridging Barriers to Entry into OT Security in the South West: Bootcamps and Tools; Grant number: 10143981. 

\bibliographystyle{IEEEtran}
\bibliography{references}

\begin{IEEEbiography}{Awais Rashid}{\,} is the architect and lead developer of LINICS at Hacktonics Ltd. Contact him at awais@hacktonics.io.
\end{IEEEbiography}

\begin{IEEEbiography}{Joseph Gardiner}{\,} is a developer of LINICS and lead developer of the ZathrICS tool at Hacktonics Ltd.  Contact him at joe@hacktonics.io.
\end{IEEEbiography}

\begin{IEEEbiography}{Louise Evans}{\,} leads user engagement and feedback for LINICS at Hacktonics Ltd.  Contact her at louise@hacktonics.io.
\end{IEEEbiography}

\end{document}